Antiferromagnetic coupling across silicon regulated by tunneling currents

R.R. Gareev[a)], M. Schmid, J. Vancea, and C. Back

*Institute of Experimental and Applied Physics, University of Regensburg, 93040 Regensburg, Germany*

R. Schreiber, D. Bürgler, and C. M. Schneider

*Institute of Solid State Research (IFF), Research Center Jülich, 52428 Juelich, Germany*

F. Stromberg, and H. Wende

*Faculty of Physics and Center for Nanointegration Duisburg-Essen (CeNIDE), University of Duisburg-Essen, 47048 Duisburg, Germany*

We report on the enhancement of antiferromagnetic coupling in epitaxial Fe/Si/Fe structures by voltage-driven spin-polarized tunneling currents. Using the ballistic electron magnetic microscopy we established that the hot-electron collector current reflects magnetization alignment and the magnetocurrent exceeds 200% at room temperature. The saturation magnetic field for the collector current corresponding to the parallel alignment of magnetizations rises up with the tunneling current, thus demonstrating stabilization of the antiparallel alignment and increasing antiferromagnetic coupling. We connect the enhancement of antiferromagnetic coupling with local dynamic spin torques mediated by tunneling electrons.

Antiferromagnetic (AF) interlayer exchange coupling between magnetic layers separated by metallic or insulating spacers is a basic effect for spintronics[1,2]. It is well established that AF coupling across metals is accompanied by the giant magnetoresistance effect, where resistance depends on the relative alignment of magnetic electrodes due to spin-dependent interface scattering[3]. Ferromagnetic layers separated by a thin tunneling barrier (TB) demonstrate interlayer coupling mediated by spin-polarized conduction electrons *via* equilibrium spin-torques[4,5]. A possible way to regulate AF coupling across TB`s[6-8] could be in utilization of dynamic spin-transfer torques produced by voltage-driven tunneling currents. Spin currents with sufficient current densities can influence total in-plane spin torques and, consequently, the strength of AF coupling[9]. This approach promises voltage-controlled AF coupling for spintronics devices.

Tunneling structures based on Si are attractive due to their compatibility with existing semiconductor technologies and prospective for spin-injection devices due to low spin-orbit scattering in this material[10]. Earlier we found strong AF coupling across Si TB`s[7]. Theoretical modeling showed that resonant impurity states in a spacer layer can lead to specific tunneling magnetoresistance (TMR) and enhanced AF coupling[11]. In accordance, we confirmed formation of a TB with a low resistance-area product[8] as well as resonant-type TMR in Si-based tunneling structures[12]. However, room-temperature (RT) spin-dependent magnetotransport in Fe/Si/Fe was not observed so far.

For our studies of RT magnetotransport across silicon TB`s and voltage-regulated AF coupling we utilized ballistic electron magnetic microscopy (BEMM). This non-destroying method with nanometer resolution was adopted earlier for studies of spin-dependent magnetocurrents in spin valves[13]. In BEMM experiments a first magnetic layer serves as a spin-polarizer and a second one - as an analyzer of magnetization alignment. Ballistic hot electrons are injected across a vacuum tunnel barrier, which prevents from leakage currents and, thus, enables measurements at RT. The energy filtering by a Schottky barrier formed close to n-doped GaAs substrate enables to separate the tunneling current $I_T$ from the ballistic collector current $I_c$. This affords a possibility to utilize tunneling currents for manipulating magnetization alignment and collector currents for detection of magnetization alignment.

We prepared AF-coupled epitaxial Fe(3nm)/Si(2.4nm)/Fe(3nm)/Au(6nm) structures on a 7-μm- thick $Ga_{0.67}P_{0.33}As$ (100) layer (n=5*10$^{16}$ cm$^{-3}$) grown on a GaAs (100) (n=1*10$^{18}$ cm$^{-3}$) wafer using thermal electron-gun evaporation as described elsewhere[7]. Before the evaporation we performed annealing of the substrate at T=870K with parallel 500 eV Ar$^+$ pre-sputtering for 0.5 hour. Epitaxial growth was confirmed from the high-energy electron emission diffraction (RHEED) GaAs-(4x6) surface reconstruction pattern. After this procedure no contamination of the surface with oxygen and carbon was detected from the Auger spectra. A detailed description of the BEMM experimental set-up is presented in Ref. 13.

The collector current $I_c$ versus biasing voltage $U_{bias}$ dependence taken at RT is demonstrated in Fig. 1. It is seen that for negative $U_{bias}$ $I_c$ starts to increase due to presence of a highly resistive Schottky barrier at the Fe/GaPAs interface (see Inset in Fig.1). For negative biasing voltages exceeding 1V the collector current depends on in-plane magnetic field H and reaches approximately 140 fA and 40 fA for H=1kOe and H=0, accordingly. Thus, we managed to detect RT magnetocurrents across TB`s based on Si.

In order to check whether the magnetocurrent really reflects the magnetization alignment we performed comparative studies of magnetotransport and magnetic properties of our Fe/Si/Fe structures. In Fig. 2 we present magnetocurrent and magneto-optical Kerr effect (MOKE) hysteresis data taken for magnetic field applied along an easy axis. It is seen that RT hysteresis in magnetocurrent correlates with the MOKE data and changes in the collector current reflect magnetization alignment. The magnetocurrent (MC) was determined from the equation: MC = [($I_{cP}$-$I_{cAP}$)/ ($I_{cAP}$)]*100% where $I_{cP}$ and $I_{cAP}$ are values of the collector current corresponding to a parallel (P) and an antiparallel (AP) alignment of magnetizations, accordingly. In our experiments the collector current is higher for the P state compared to the AP state and exceeds 200% for $I_T$=30 nA. For higher tunneling currents the magnetocurrent is even bigger (see Fig. 3). The same sign of magnetocurrent was found for spin valves and explained by spin-filtering effects due to increased scattering for minority-spin channel[13].

Next we studied the magnetocurrent for different values of $I_c$. The increase of the saturation field $H_{sat}$ with the tunneling current $I_T$ (Fig. 3) upon reorientation of magnetizations between AP and P states, indicates an enhancement of AF coupling by increasing $I_T$. Actually, as it is seen in Fig. 3 the saturation field increases from $H_{sat}$~300 Oe to $H_{sat}$ ~900 Oe upon rising of $I_T$ from 30nA to 50nA. For $I_T$=60 nA magnetic field H= 1000 Oe is even not sufficient to switch magnetization from the AP state. For $I_T$ =30 nA and below the tunneling current does not influence $H_{sat}$ and AF coupling, accordingly. We determined the strength of AF coupling by fitting MOKE hysteresis curves as described elsewhere[7] and found /$J_1$/~35 µJ/m$^2$ in this regime. For higher $I_T$ the $H_{sat}$ increases and transition from AP to P state becomes sharply pronounced. Taking into account that the bilinear coupling term $J_1$ is proportional to $H_{sat}$[14] we estimated that AF coupling reaches /$J_1$/ ~110 µJ/m$^2$ ($I_T$=50 nA) and, finally, /$J_1$/ is exceeding 120 µJ/m$^2$ for $I_T$=60 nA. Thus, by increasing twice the tunneling current AF coupling increases by a factor of three at least. It is interesting that that for $I_T$=40nA the collector current shows a hysteretic behavior probably due to formation of a multi-domain structure. Oscillations of the collector current in the magnetic field (Fig. 3) could be related to RT charging effects[15] upon resonant tunneling across nanometer-scaled impurities. A detailed study of charging effects is in progress and out of scope of this report.

We explain the enhancement of AF coupling by dynamic local spin torques due to voltage-assisted tunneling currents. Actually, for ballistic tunneling currents the current density j becomes sufficient for producing substantial local dynamic spin-transfer torques. In the AP state spin-polarized electrons from the upper iron layer exert dynamic torques in the bottom iron layer. These dynamic spin torques give rise to additional pinning of the bottom iron layer, which can lead to switching of the upper iron layer instead of coherent rotation of both magnetic layers in magnetic field as observed in our experiments for $I_T$ above 30 nA. Assuming that tunneling current flows from the last atom of the tip (diameter of the current spot near 0.3 nm) we obtain for $I_T$~50 nA current densities exceeding $5*10^7$ A/cm$^2$. These j values, as shown for nanopillars with synthetic AF free layers[16] are sufficient for switching of magnetization. Charging effects in magnetic tunnel junctions can also enhance spin torques compared to metallic multilayers[17]. It should be noted that uncompensated stationary spin torques and spin accumulation also play a decisive role in AF coupling across tunneling structures based on MgO[18]. The equilibrium spin-torque mechanism with different spin filtering at diffused interfaces is most favorably describing AF coupling in our Fe/Si/Fe structures as well. In this scenario the majority (spin-up) electrons could be strongly reflected, while minority (spin-down) electrons are more propagating. Consequently, the spin-up electrons exert spin-transfer torques to the first iron layer and spin-down electrons produce equilibrium spin torques in the opposite direction in the second iron layer, thus stabilizing AF alignment at zero biasing voltage. In presence of biasing voltage dynamic spin torques start to play role and favor stabilization of P or AP alignment depending on the direction of the tunneling current as confirmed earlier by dynamic stability diagram for spin-valves in the non-precessional regime[19]. In our experiments dynamic spin torques favor antiparallel alignment of magnetizations. Finally, for comparable dynamic and equilibrium spin torques an increased AF coupling assisted by tunneling spin-polarized currents becomes detectable.

Concluding, we found RT ballistic magnetocurrent in AF-coupled Fe/Si/Fe structures. The hot-electron collector current reflects alignment of magnetizations with the magnetocurrent exceeding 200% at RT. The saturation magnetic field corresponding to the collector current in the parallel alignment of magnetizations increases with tunneling current, thus demonstrating stabilization of the antiparallel alignment and increased AF coupling. We connect the enhancement of AF coupling with local dynamic spin torques mediated by voltage-driven spin-polarized tunneling electrons.

This work is supported by the Project DFG 9209379.

Figure captions:

FIG. 1. The collector current $I_c$ *versus* biasing voltage $U_{bias}$ taken in the remanent state (H=0: quadrats) and above the saturation field (H=1 kOe: circles) applied along easy axis [110] direction. The inset demonstrates the resistance R *versus* $U_{bias}$ for the Fe/Ga$_{0.67}$P$_{0.33}$As Schottky barrier.

FIG. 2. Longitudinal MOKE hysteresis and collector current $I_c$ hysteresis taken at $U_{bias}$ =-2,5V. The in-plane magnetic field is aligned along the easy-axis [110] direction. The $I_c(H)$ hysteresis loops are averaged over 20 cycles. Thin arrows indicate the direction of the field, thick arrows show magnetization alignment.

FIG. 3. Single-cycle collector current $I_c$ *versus* magnetic field aligned along the easy-axis [110] direction for different values of the tunneling current $I_T$ measured at biasing voltage U= -2.5V. The quadrats and circles correspond to positive and negative magnetic field sweep directions indicated by arrows, correspondingly. Parallel and antiparallel alignments of magnetizations are marked as P and AP, accordingly.

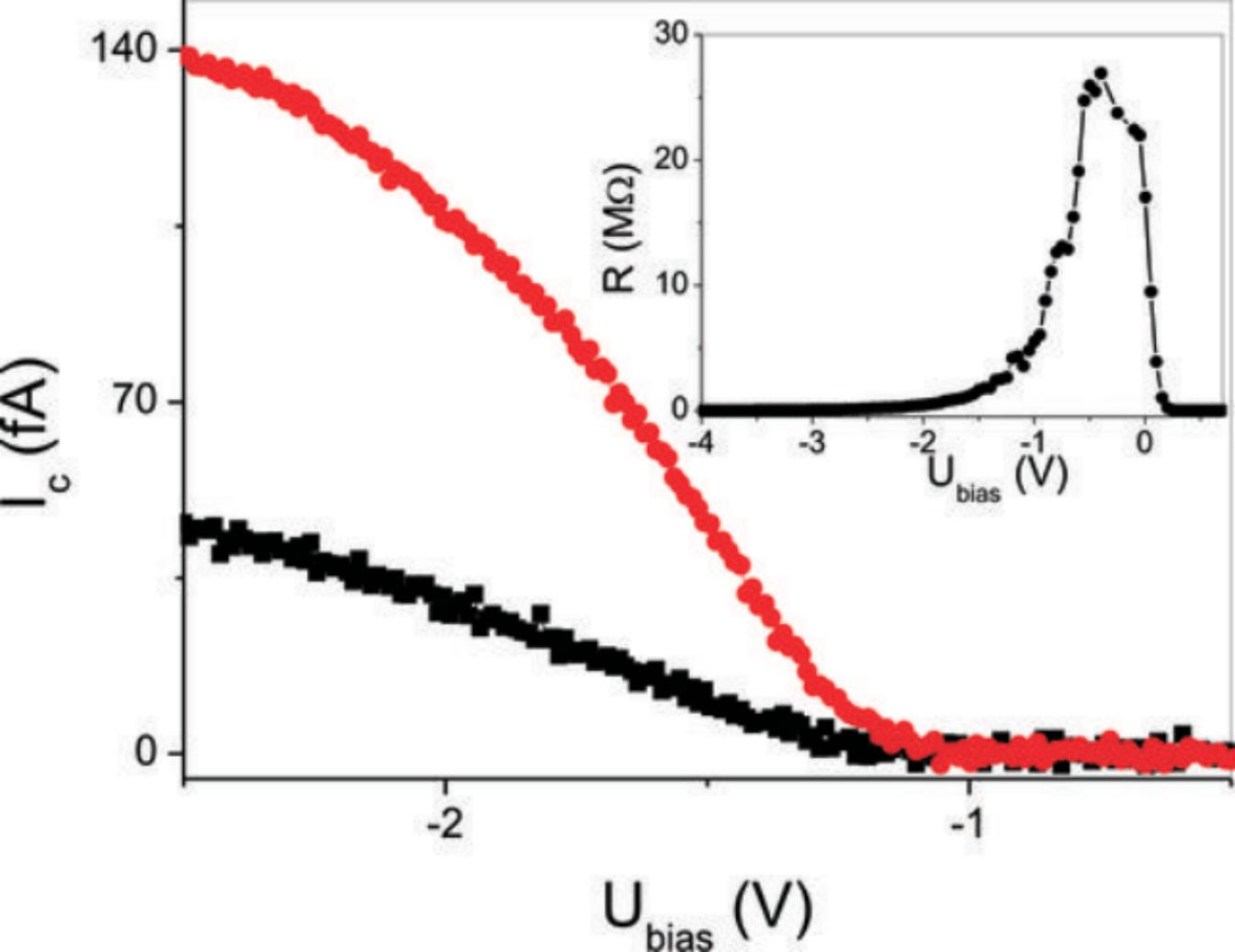

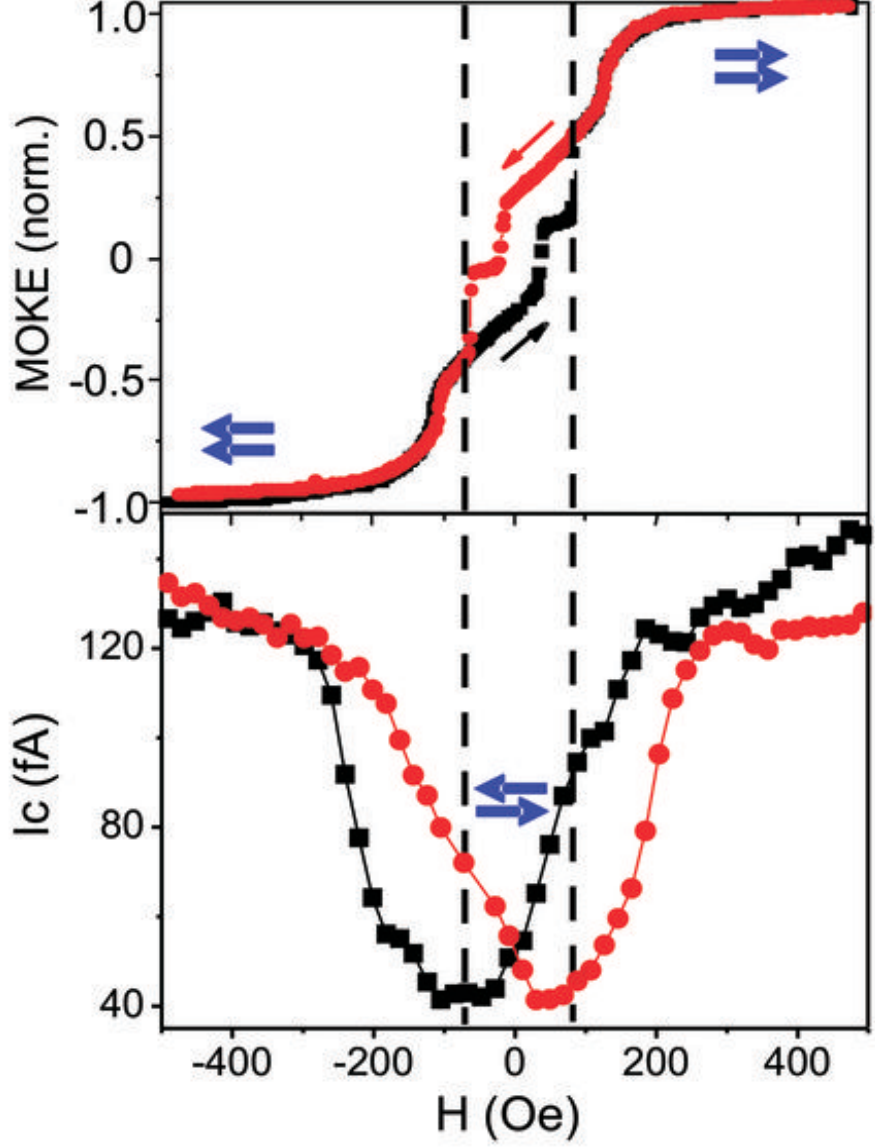

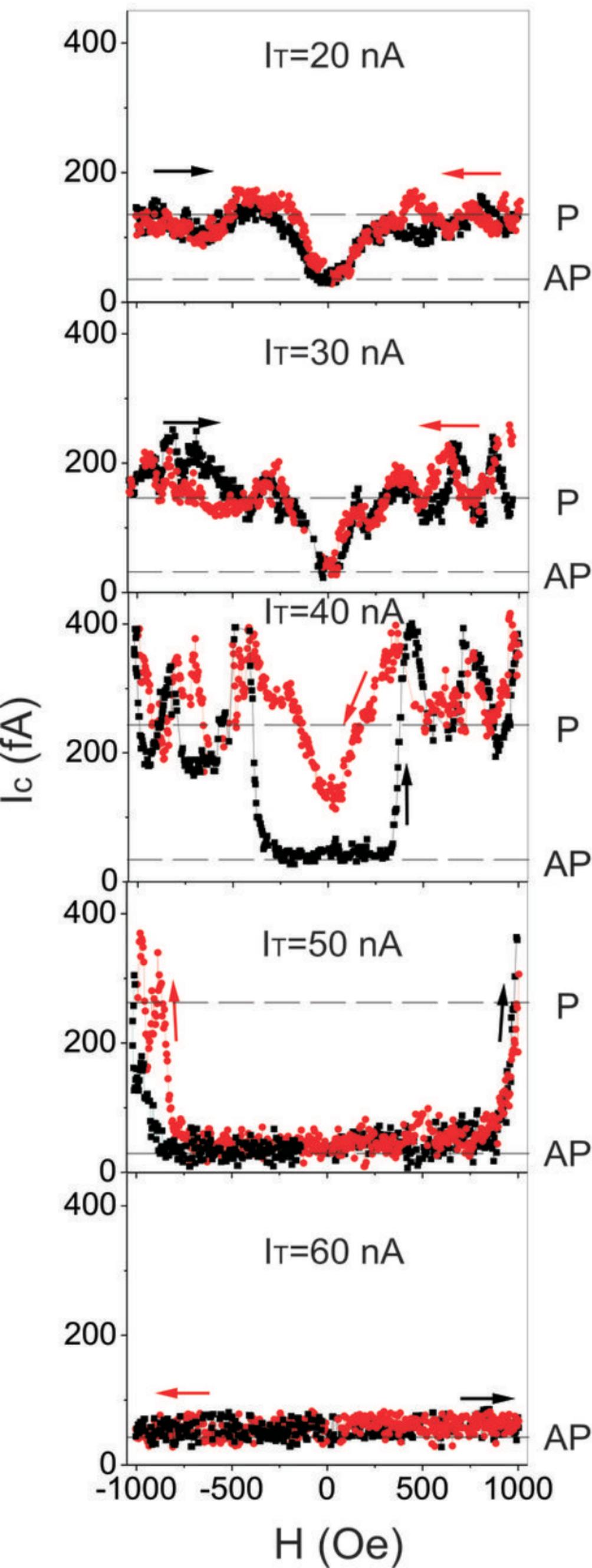